\documentclass[reprint,amsmath,amssymb,floatfix,prl,aps,superscriptaddress]{revtex4-1}

\usepackage{graphicx}
\usepackage{amsmath}
\usepackage{amssymb}
\usepackage{upgreek}
\usepackage{subfigure}
\usepackage{bbold}

\def\br{{\mathbf{r}}}

\def\Rb87{^{87}\text{Rb}}                     
\def\Na23{^{23}\text{Na}}                     
\def\Li6{^{6}\text{Li}}                       
\def\muB{\mu_{\rm B}}         

\def\bra#1{\mathinner{\langle{#1}|}}
\def\ket#1{\mathinner{|{#1}\rangle}}

{\catcode`\|=\active
  \gdef\Braket#1{\left<\mathcode`\|"8000\let|\BraVert {#1}\right>}}
\def\BraVert{\egroup\,\mid@vertical\,\bgroup}

\newcommand{\ex}{\mathbf{e}_x}
\newcommand{\ey}{\mathbf{e}_y}
\newcommand{\ez}{\mathbf{e}_z}
\newcommand{\benum}{\begin{enumerate}}
\newcommand{\eenum}{\end{enumerate}}

\begin{document}

\title{Magnetic-field-mediated coupling and control in hybrid atomic-nanomechanical  systems}
\author{A.~Tretiakov}
\affiliation{Department of Physics, University of Alberta, Edmonton AB, Canada}
\author{L.~J.~LeBlanc}
\affiliation{Department of Physics, University of Alberta, Edmonton AB, Canada}
\affiliation{Canadian Institute for Advanced Research, Toronto, ON, Canada}

\begin{abstract}
Magnetically coupled hybrid quantum systems enable robust quantum state control through Landau-Zener transitions.  Here, we show that an ultracold atomic sample magnetically coupled to a nanomechanical resonator can be used to cool the resonator's mechanical motion, to measure the mechanical temperature, and to enable entanglement of more than one of these mesoscopic objects.  We calculate the expected coupling for both permanent-magnet and current-conducting nanostring resonators and describe how this hybridization is attainable using recently developed fabrication techniques, including SiN nanostrings and atom chips.

\end{abstract}

\maketitle


Hybrid quantum systems serve to bring the advantages of multiple quantum technologies together~\cite{Treutlein:2014go,Kurizki:2015ew}.  Currently, no individual platform is ideal -- all systems have advantages for performing some tasks, while retaining less-desirable properties in other realms.  Whether it is coherence time, facility for exchanging quantum information, or data processing speeds, the reason for an advantage is usually fundamentally tied to a difficulty. For ultracold atoms,  isolation from the environment enables excellent coherence and state control, but hinders information transmission to conventional read-out technologies. Combining platforms to exploit the advantages and make irrelevant the disadvantages can lead the way to viable hybrid quantum technologies.

Along with ultracold atoms, nanomechanical solid-state devices are among the leading candidates as components of hybrid systems~\cite{Greenberg:2012gi,Poot:2012fha,Aspelmeyer:2013vr}; unlike quantum gases, they have good readout but poor coherence times. Quantum correlations can be transferred between these very different platforms using electric and magnetic field couplings. Ultracold atoms and solid state device hybridization was demonstrated in a variety of systems: with optical fields to cavity modes~\cite{Ye:1999vj,Raimond:2001jj} or vibrating membranes~\cite{Camerer:2011do,Vogell:2013fr,Jockel:2014cz}, and with magnetic fields via nanomechanical magnetic resonators~\cite{Hunger:2011eo,Wang:2006dd,Montoya:2015el}.  As we will explain in detail, coherent  coupling between the atomic and mechanical systems can be used   to cool the mechanical motion of the resonators, to measure the mechanical system's temperature, and to transfer correlations between mechanical devices to realize entanglement.

Here, we focus on systems where oscillating magnetic fields  are used to drive transitions between long-lived ground states.  While these transitions are inherently weaker than those from optical fields, the states' lifetimes are appealing for coherent quantum state control and reversible transfer of quantum coherence between systems.  Furthermore, temporal ``Landau-Zener'' sweeps of an external magnetic field~\cite{Rubbmark:1981dg,OKeeffe:2013ij} can be used to flip atomic spins while changing the phonon occupation in coupled systems. Unlike  previously implemented cantilever designs that have been used to couple to ultracold atoms,  we consider here SiN nanostring resonators~\cite{Verbridge:2006ix,Verbridge:2008ja,Biswas:2012dba,Darazs:2014gd} fabricated with high-tensile stress.  For these devices, the mechanical behaviour is dominated by the stress in the string~\cite{Schmid:2011hi} and not the material properties, ultimately leading to high quality factors $Q$ for the resonators, even at room-temperature.



\begin{figure}[t!]  
\begin{center}
\includegraphics[width=85mm]{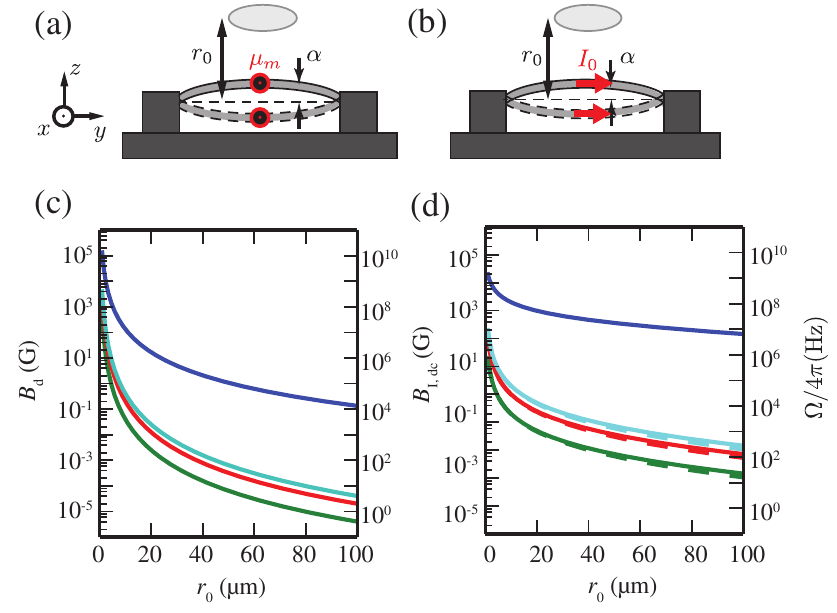} 
\caption{Nanomechanical magnetic resonator schematics with (a) a permanent magnetic dipole, $\mu_m$ and (b) a current carrying wire. Calculated magnetic field (left axes) and coupling parameters for $^{87}$Rb atoms with $F=1$ (right axes) from (c) a nanostring with a permanent magnet with $\mu_m =  0.067$~nJ/T  and (d)  a current-carrying wire (calculated to give the per-Ampere coupling). The blue curves indicate the magnitude of the time-independent field component  $B_{0z,{\rm d}}$, while other colours show time-dependent parts $b_{\rm d} \alpha$ for amplitudes $\alpha = $ 10 nm (light blue), 5 nm (red), and 1 nm (green). Dashed curves show numerical comparisons.
} 
\label{fig:fig1}
\end{center}
\end{figure}

In this work, we frame our discussion in terms of the magnetic-field mediated coupling between neutral alkali metal atoms and a nanomechanical resonator whose magnetic field is generated either by a current~\cite{Biswas:2012dba} or by a permanent magnet~\cite{Diao:2013ca} (Fig.~\ref{fig:fig1}a and b.)  In a magnetic field, the Zeeman effect lifts the degeneracy within the ground states,  which we describe using the generalized spin operator $\mathbf{\hat{F}} = (\hat{F}_x,\hat{F}_y,\hat{F}_z)$. For the ground states $\ket{F,m_F}$, the Zeeman Hamiltonian is $\hat{H}_{\rm Z}  = - \boldsymbol{\hat{\mu}}\cdot \mathbf{B} = g_F \mu_{\rm B}( \mathbf{ \hat{F}}\cdot \mathbf{B})/\hbar$,
where $g_F$ is the Land\'e $g$-factor and $\mu_{\rm B}$ is the Bohr magneton, and we assume the weak-field limit of the Zeeman effect.  
Near a nanomechanical magnetic resonator with resonant frequency $\omega_{\rm m}$, the associated oscillating magnetic field $\mathbf{B}_{\rm m}(\br,t) = \mathbf{B}_{\rm m, 0}(\br) \cos\omega_{\rm m} t$  drives transitions between ground states. In a static magnetic field,  the oscillating field couples to the electronic spin, and the magnetic dipole operator is $\hat{\boldsymbol{\mu}} = -g_F\muB \hat{\mathbf{F}}/\hbar$.  For levels $\ket{F m_F}$ and $\ket{F^\prime m_F^\prime}$  separated by the atomic energy level spacing $\hbar \omega_{\rm a}$ due to a static magnetic field along $\ez$, an oscillating magnetic field with amplitude $ \mathbf{B}_{\rm m, 0}  =  {B}_{\rm m, 0} \{\sin\theta, 0, \cos\theta\}$ and frequency $\omega_{\rm m} = \omega_{\rm a}$ results in the resonant  coupling element 
\begin{align}
\hbar\Omega = \frac{  g_F \muB{B}_{\rm m, 0}}{\hbar}\bra{F^\prime m_F^\prime} \hat{F}_x \sin\theta+\hat{F}_z \cos\theta \ket{F m_F}.
\label{eq:Omega}
\end{align}   
The first term in the matrix element  corresponds to transitions with $m_F^\prime=m_F\pm1$. For two levels with the same $F$, in the rotating-wave approximation, the coupling parameter is\begin{align}
\frac{\hbar\Omega}{2}=\frac{g_F \mu_{\rm B} B_{\rm m, 0}}{4}\sqrt{F(F+1)-m_F(m_F\pm1)}.
\label{eq:Rabi}
\end{align} 
The second term in Eq.~\ref{eq:Omega} describes transitions that preserve $m_F$ but change $F$, corresponding to transitions between hyperfine levels, which are allowed, but are typically at significantly higher frequencies, and are not discussed here.

One method for creating magnetic coupling  between a nanostring and atoms  is to deposit a permanent magnet on a small region of the nanostring~\cite{Diao:2013ca} (Fig.~\ref{fig:fig1}a).  If the permanent magnet is treated as a point-like  dipole $\boldsymbol{\mu}_m$ oriented along $\ex$, the field created  at a distance $r_0$ above the centre of the nanostring vibrating along $\ez$ with an amplitude $\alpha$ is~\cite{Steinke:2011ig}
\begin{align}
{B}_{\rm d}(t) = \frac{\mu_0 \mu_m}{4\pi r_0^4}[r_0-3 \alpha \cos(\omega_{\rm m} t)] \nonumber \\
\equiv B_{0x,{\rm d}} + b_{\rm d} \alpha \cos(\omega_{\rm m} t).
\label{eq:dipole}
\end{align}
Here, we have separated the magnetic field into stationary $ B_{0,x{\rm d}}$ and  time-dependent $b_{\rm d} \alpha$ amplitudes, where the gradient of the field $b_{\rm d} = 3 \mu_m\mu_0/(4\pi r_0^4)$  determines the strength of the spin-flipping field component. 
In general, the static part is  a few orders of magnitude larger than oscillatory part (see Fig.~\ref{fig:fig1}c), which leads to additional challenges in experimental design~\cite{Note1}.

A second method for implementing magnetic coupling between a nanostring and atoms is to pass current through a conductive nanostring~\cite{Sulkko:2010fc,Larsen:2011ji,Kalman:2012ko}.  The magnetic field created by a long wire carrying current $I_0$ along $\ey$ and vibrating along $\ez$ is
\begin{align}
{B}_{\rm I, dc}(t) = \frac{\mu_0 I_0}{2\pi r_0^2}[r_0-{\alpha}\cos(\omega_{\rm m} t)]\nonumber\\
  \equiv B_{0x,I} + b_{\rm I} \alpha \cos(\omega_{\rm m} t).
\label{eq:DC_field}
\end{align}
Here, the spin-flipping amplitude $b_I$ scales with  distance as  $1/r_0^2$, which differs from the permanent magnetic dipole moment where $b_{\rm d} \propto 1/r_0^4$. Compared to cantilever geometries,  nanostrings facilitate conduction and can take advantage of this improved scaling.  Comparisons of this analytical model (Eq.~\ref{eq:DC_field}) to a numerical simulation show good agreement (see Fig.~\ref{fig:fig1})
~\cite{Note2}.

If an alternating, out-of-phase current $I_{\rm ac} = I_0 \sin \omega_{\rm dr} t$ passes through the nanostring and $\omega_{\rm ac}$ differs from the atomic resonance frequency,  the atoms experience a field 
\begin{align}
{B}_{\rm I,ac}(t)= \frac{\mu_0 I}{2\pi r_0^2}[r_0\sin\omega_{\rm ac} t
  +\frac{\alpha}{2}\sin(\omega_{\rm ac}+\omega_{\rm m}) t \nonumber\\
 + \frac{\alpha}{2}\sin(\omega_{\rm ac}-\omega_{\rm m}) t ].
\end{align}
The first term arises from the alternating current itself, while the second  and third are due to motion of the resonator.
If the wire is driven into mechanical oscillation at  $\omega_{\rm dr}$ and resonant with the atomic sample at $\omega_{\rm dr}+\omega_{\rm m} = \omega_{\rm a}$, only terms proportional to the mechanical amplitude $\alpha$ address the atoms.  This separation of frequencies differentiates transitions driven by mechanical motion from transitions due to fields generated by currents in other parts of the device~\cite{Wang:2006dd}.  

Cryogenically cooling the mechanics can  reduce energy scales to the level at which  quantization of mechanical motion matters~\cite{Teufel:2011jg,Meenehan:2015ie,Wang:2015de,Riedinger:2016cl}. To describe this, we promote the resonator's displacement to an operator: $\alpha \rightarrow \alpha_0(\hat{a} + \hat{a}^\dagger)$, where $\alpha_0 = \sqrt{\hbar/2m_{\rm eff} \omega_{\rm m}}$ is the zero-point motion of the resonator, $m_{\rm eff}$ is its effective mass, and  $\hat{a}$ and $\hat{a}^\dagger$  are mechanical-mode phonon annihilation  and creation operators of the mechanical system. By isolating  two levels in the atomic system, where for example $\ket{\uparrow}$ and $\ket{\downarrow}$ represent two states with $m_{F\uparrow}$ and $m_{F\downarrow}$, the operator $\hat{F}_x$ becomes $\hat{\sigma}_x = \hat{\sigma}^+ + \hat{\sigma}^-$. 
In the rotating-wave approximation, this yields a Hamiltonian analogous to the Jaynes-Cummings model~\cite{Shore:1993gca}
\begin{align}
\hat{H}_{\rm at} =  \hbar \omega_{\rm m} \hat{a}^{\dagger} \hat{a} +\frac{\hbar \omega_{\rm a}}{2} \hat\sigma_z + \frac{\hbar g_0}{2} (\hat\sigma^+\hat{a} +\hat\sigma^-\hat{a}^\dagger),\label{eq:JCH}
\end{align}
where $\hat\sigma_z = \hat\sigma^+\hat\sigma^- - \hat\sigma^-\hat\sigma^+ $, 
and $\hbar g_0= g_F \muB b \alpha_0[F(F+1) - m_{F\uparrow}m_{F\downarrow}]^{1/2}/2$ is the single-atom-single-phonon coupling parameter.  

One of the great advantages of hybridizing systems with ultracold atoms is the ability to couple single excitations collectively to an ensemble of $N$ atoms~\cite{Shore:1993gc,TreutleinPRL2007,Verdu:2009kt}, such that the collective spin operators are $\hat{\tilde\sigma}^+ = (1/\sqrt{N})\sum_i \hat\sigma^+_i$ and $\hat{\tilde\sigma}^- = (1/\sqrt{N})\sum_i \hat\sigma^-_i$, where the sums run over all $N$ atoms. This many-atom system has an enhanced effective coupling parameter: $g_{\rm eff} = \sqrt{N} g_0$, and Eq.~\ref{eq:JCH} is transformed to the Tavis-Cummings Hamiltonian~\cite{Shore:1993gca}.	To generalize, we simplify notation for the coupled system:  $\ket{n,m_F}\rightarrow \ket{n,s}$ where $n$ is the number of phonons excited in the resonator and $s = (\uparrow, \downarrow)$ is the spin-state of the two-level atomic system.

\begin{figure}[t!]
\begin{center}
\includegraphics[width = 85mm]{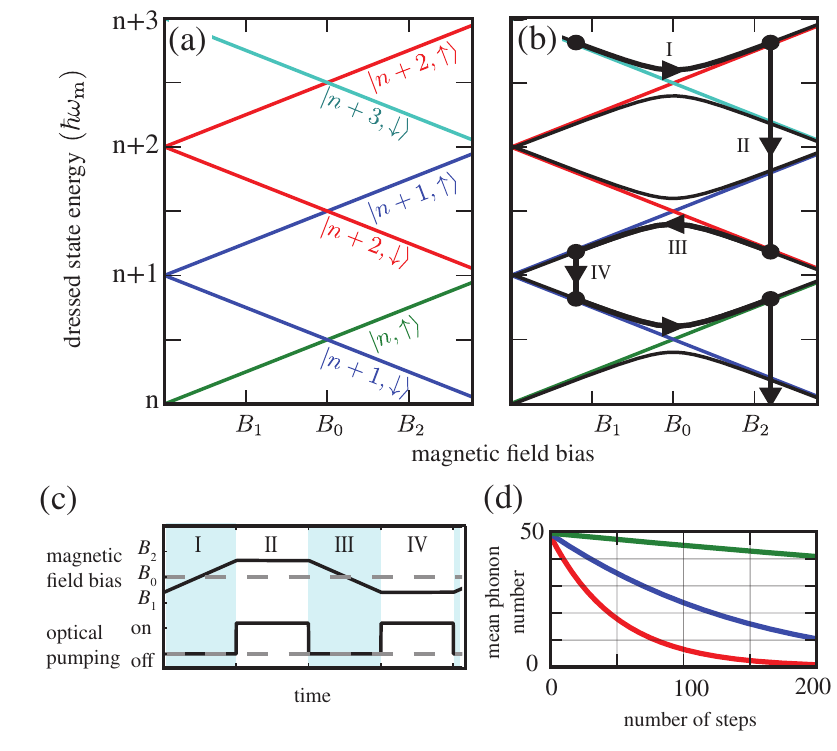}
\end{center}
\caption{Mechanical cooling scheme. (a) Ladder of uncoupled phonon-spin dressed states. (b) Coupled phonon-spin states (black), with thick lines indicating magnetic field sweeps I, III  and V, and  optical-pumping spin-flip purification II and IV. (c) Conceptual timing sequence for magnetic field sweeps (blue shaded regions) and optical pumping (OP) pulses (white regions) I through V. (d) Reduction in mean phonon number, per spin, as the cooling sequence progresses, for an initial $\bar n = 50$ and $p_{\rm LZ}(1) = 1$ (red), 0.5 (blue), and 0.1 (green).  Each ``step''  is one sweep and one OP pulse.}
\label{Fig:rachet}
\end{figure}

Quantum state manipulation becomes possible when control over both the resonator's phonon occupation and the atoms' spin (either collective or  single-atom) are possible. Consider a basis of  single-spin-atomic states dressed by the phonon-occupation states, where the dressed-state energies are  magnetic-field dependent,  as shown in Fig.~\ref{Fig:rachet}(a).  In the presence of coupling between the mechanical mode and the atomic spin,  magnetic field values $B_0$ for which $\ket{n+1,\downarrow}$ cross $\ket{n,\uparrow}$ are avoided, and coupled eigenstates [Fig.~\ref{Fig:rachet}(b)] connect different phonon occupations.  

With this connection, we introduce a scheme that uses magnetic-field sweeps across  resonance to manipulate quantum states. An adiabatic magnetic field sweep from $B_1< B_0$ to $B_2>B_0$  transfers atoms from one spin state to the other as it transfers the resonator from one phonon-number occupation state to a neighboring one.  Landau-Zener (LZ) theory~\cite{Rubbmark:1981dg,OKeeffe:2013ij} gives the probability of  complete state transfer 
\begin{align}
p_{\rm LZ}(n)=1- \exp\left[ -\frac{\pi\hbar ng_0^2}{2g_F\mu_{\rm B}(\Delta B/\Delta t)}\right],
\end{align}
where we assume a  linear sweep of the magnetic field $\Delta B$ in a time $\Delta t$ (which is sufficiently long).  Here, $g_0$ refers to either the single atom or collective coupling, as the situation dictates, and the $n$-dependence reflects the amplitude dependence of the coupling, as is well-known in a Jaynes-Cummings ladder.  In general, states  transform from $\ket{n,\uparrow} \rightarrow [\sqrt{p_{\rm LZ}}\ket{n+1,\downarrow}+\sqrt{1-p_{\rm LZ}}e^{i\phi}\ket{n,\uparrow}] $ or $\ket{n,\downarrow} \rightarrow [\sqrt{p_{\rm LZ}}\ket{n-1,\uparrow}+\sqrt{1-p_{\rm LZ}}e^{-i\phi}\ket{n,\downarrow}]$. By varying the coupling strength and field-sweep duration, we can control  $p_{\rm LZ}$ and $\phi$ to create different states of the system. Efficient state transfer is limited by the available coupling $g_0$, though the collective-spin $\sqrt{N}$-enhancement of $g_0 \rightarrow g_{\rm eff}$ and the ability to bring atoms into close proximity to the resonator with atom-chip trapping techniques facilitates this control over the quantum states.

A significant  advantage to hybridizing mechanical devices with atoms is the wide range of opportunities in applications for quantum state control.
Here we discuss cooling a resonator's mechanical motion, nanomechanical thermometry, and entangling multiple resonators.  In all cases, these protocols are robust to technical noise that can  arise due to magnetic field fluctuations, so long as the sweep's beginning and ending points are far from resonance.

The  proficiency with which atoms can shed energy though the emission of electromagnetic radiation enables the cooling of coupled mechanical systems. In previous work, optical coupling between atoms and a membrane cooled  mechanical motion~\cite{Hammerer:2010fq,Jockel:2014cz}. Here, we propose a scheme to remove phonon excitations from the mechanics by selectively flipping the spins of the atomic system using a series of magnetic field sweeps and optical pumping pulses (Fig.~\ref{Fig:rachet}).  We describe the process for a single spin, but imagine the process proceeding in parallel for many spins.  To begin, the spin must be optically pumped into the low-field seeking state $\ket{\downarrow}$.  Next, the sequence of sweeps and pulses begins.  First, a magnetic field sweeps across the resonance from $B_1$ to $B_2$ to transfer the atom to the high-field seeking state $\ket{\uparrow}$, and onto the branch with one less phonon (I). Second, an optical pumping pulse resonant at this higher field $B_2$  flips any $\ket{\uparrow}$ spins  to $\ket{\downarrow}$ (II).  Third, the magnetic field sweeps back from $B_2$ to $B_1$ to put the spin into $\ket{\uparrow}$, on the branch with one less phonon (III).  Fourth, an optical pumping pulse resonant at $B_1$ flips $\ket{\uparrow}$  spin to $\ket{\downarrow}$ (IV).  From this point, the sequence repeats, and the number of phonons in the mechanical resonator is sequentially reduced.  The use of optical pumping pulses serves to ``reset'' the system after each LZ sweep, and ensures that spins do not get on the wrong branch, upon which they would act heat the resonator. Furthermore, the spontaneous emission associated with these pulses carries entropy out of the system.

Even if the probability $p_{\rm LZ}$ of a spin-flipping transition for each individual process is small, that fraction of an ensemble is active in the cooling process. In the high-temperature limit ($\bar{n} \gg 1$), each sweep across resonance with LZ probability $p_{\rm LZ}$  in a system with $N$ atomic spins and a mean number $\bar n$ of mechanical phonons removes as many phonons as atoms that undergo a spin flip: $\Delta \bar n = p_{\rm LZ}(\bar n) N$, and each sweep (assuming no thermalization) yields $\Delta T = \hbar \omega_{\rm m} N \Delta \bar{n}/k_{\rm B}$, where this temperature is associated with the phonon occupation of mechanical  modes. For  $\omega_{\rm m}/2\pi = 850~{\rm kHz}$ and  $N = 10^5$, this change in temperature is $\Delta T = p_{\rm LZ}(\bar n) \times 4.1$~K for each sweep.  Though this cooling may be modest for room-temperature systems, this scheme does not lose efficacy until the $n=0$ mode is significantly occupied, as seen in the calculation of the phonon number reduction in Fig.~\ref{Fig:rachet}d (which does not assume the high-temperature limit). This type of cooling should have  a significant impact on cryogenically cooled systems requiring the transition from 4~K to lower temperatures.   For a single  $^{87}$Rb atom in its $F = 1$ ground state and a  nanostring frequency with $\omega_{\rm m}/2\pi = 850$~kHz  ($B_0= 1.43$~T) and $m_{\rm eff}=8.4\times 10^{-13}$~kg, if the atom is a distance $r_0 = 1~\mu$m, we find  $g_0 =  21$~s$^{-1}$ and magnetic field ramps over 1 mG must last about 1 s to provide  efficient of cooling at the lowest phonon occupations~\cite{Note1}, though this time scales as $1/n$ and these times may be faster for higher phonon occupations. Further,  cooling must happen faster than rethermalization in order to be effective, which is viable for the high-$Q$, high-frequency nanostrings  we consider here ($Q \approx 10^6$)~\cite{Verbridge:2008ja,Biswas:2012dba}.

\begin{figure}[t!]
\begin{center}
\includegraphics[width = 85mm]{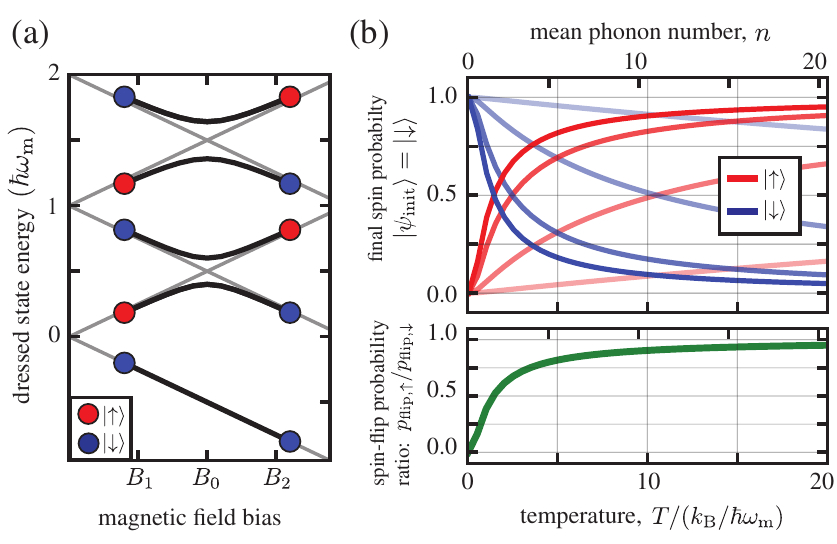}
\end{center}
\caption{An atomic thermometer for mechanics.  (a) Near the bottom of the phonon-occupation ladder, the probability of spin-state transfer from $\ket{\uparrow} \rightarrow \ket{\downarrow}$ depends on the LZ probability $p_{\rm LZ}$, except for the lowest state, for which $p_{\rm LZ}(0) = 0$.  (b, upper) Assuming Bose-Einstein statistics of phonon-number occupation in the mechanical mode, the probability of a spin-flip from  $\ket{\downarrow}$ depends on temperature for all $p_{\rm LZ}$ (darkest curve is $p_{\rm LZ}(1) = 1.0$, with $p_{\rm LZ}(1) = 0.75,~0.50,~0.25$ also shown with respectively fainter curves.)  (b, lower) The ratio of transfer probability $p_{\rm flip, \downarrow}$ from $\ket{\downarrow}$ to  $p_{\rm flip, \uparrow}$ from $\ket{\uparrow}$ is is independent of $p_{\rm LZ}(n)$ and depends strongly on temperature below $\bar{n}\approx10$, making this atomic-state probability measurement a good temperature indicator.}
\label{fig.thermometer}
\end{figure}

The asymmetry of state transfer at low phonon numbers~\cite{Teufel:2011jg,Weinstein:2014fc} offers a unique opportunity for mechanical thermometry.  For single sweeps across resonance, spin flip probability $p_{\rm flip}$ depends on the occupation of the $n = 0$ phonon mode.  Consider sweeps from high-to-low field (Fig.~\ref{fig.thermometer}a): atomic spins prepared in $\ket{\uparrow}$ (through optical pumping, for instance) will all experience spin flips with the probability $p_{\rm flip, \uparrow} = p_{\rm LZ}$, while atomic spins prepared in $\ket{\downarrow}$ will undergo transfer with $p_{\rm flip, \downarrow} = p_{\rm LZ}$ only if the phonon occupation $n>0$.  Fig.~\ref{fig.thermometer}b shows the probability of spin flip in this second case, for phonon occupations at mechanical temperatures $T$ assuming Bose-Einstein statistics
~\cite{Note1}.
The ratio of spin transfer between initial states $\ket{\uparrow}$ and initial states $\ket{\downarrow}$ [measured either using several measurements in single- or low-atom number experiments, or in an ensemble measurement for large-atom-number experiments] is independent of $p_{\rm LZ}$ and depends only on the ground state population, given by the Boltzmann factor such that $p_{\rm flip \downarrow}/ p_{\rm flip \uparrow}= \exp(-\hbar \omega_{\rm m}/k_{\rm B} T)$.  As shown, this ratio of probabilities is strongly temperature-dependent below a mean phonon occupation of approximately 10, enabling sensitive thermometry in a difficult-to-access regime~\cite{Gorodetsky:2010wk,Teufel:2011jg,Meenehan:2015ie,Wang:2015de,MacDonald:2016ga,Riedinger:2016cl} using LZ sweeps.

Another avenue for quantum control using these LZ sweeps is to entangle spatially separated resonators by individually coupling to the same atomic ensemble at different times (for example, by transporting the atoms from one location to the other in a magnetic or optical trap). Consider the initial state of two Fock-like resonators $\ket{\psi_i} = \ket{n_1, n_2, \uparrow}$.  A two-step procedure starts with the atoms near the first mechanical system (with $n_1$ phonons) that first undergoes a sweep with probability $p_{\rm LZ} = p_1$.  Next, the atoms are moved to the second mechanical system (with $n_2$ phonons) and a second sweep with probability of $p_{\rm LZ} = p_2$ results in a final state of the combined system  
\begin{align}
\ket{\psi_f} = \left[\sqrt{p_1p_2} \ket{n_1+1, n_2-1,\uparrow} +\sqrt{\bar{p}_1\bar{p}_2} \ket{n_1, n_2,\uparrow}\right]  \nonumber
\\ + \left[\sqrt{p_1\bar{p}_2} \ket{n_1+1, n_2,\downarrow} +\sqrt{\bar{p}_1p_2} \ket{n_1, n_2+1,\downarrow}\right], \nonumber
\end{align}
where $\bar{p} = 1-p$. In the case where $p_1 = p_2 = 1/2$, a spin measurement  will project the two resonators in an entangled state: either  $\ket{\psi_f} = ( \ket{n_1+1, n_2} + \ket{n_1, n_2+1})/\sqrt{2}$ upon making an atomic measurement $\ket{\uparrow}$ or $\ket{\psi_f} = ( \ket{n_1, n_2} + \ket{n_1+1, n_2-1})/\sqrt{2}$ if the atomic measurement yields $\ket{\downarrow}$.   Near the ground state, when $n_1$ or $n_2 = 0$, a sweep across resonance cannot flip the spin ($\ket{0,\downarrow} \rightarrow \ket{0,\downarrow}$).  In the case where $n_1 = n_2 = 0$ and the initial state is $\ket{\psi_i} = \ket{0,0, \uparrow}$, the final state in the protocol described above results in the state $\ket{\psi_f} =  \sqrt{p_1} \ket{1,0,\downarrow} + \sqrt{\bar p_1p_2} \ket{0,1,\downarrow}+ \sqrt{\bar p_1\bar p_2} \ket{0,0,\uparrow}$.  When  $p_1 = (1-p_1)p_2$, the resonators are left in an equal-amplitude entangled state upon a $\ket\downarrow$ measurement of the atomic spin.  The special case where $p_1 = 1/2$ and $p_2 = 1$ yields the state [$\ket{\psi_f} = (\ket{1,0} + \ket{0,1})/\sqrt{2}$] in the resonator even without measuring the atoms.

Through this protocol, the atoms act as a coherent bus for quantum correlations, such that unconnected  nanomechanical devices can be entangled.  Upon expanding this protocol to many nanostrings, this can be used to create a network of coupled resonators.

\begin{figure}[t!]  
\begin{center}
\includegraphics[]{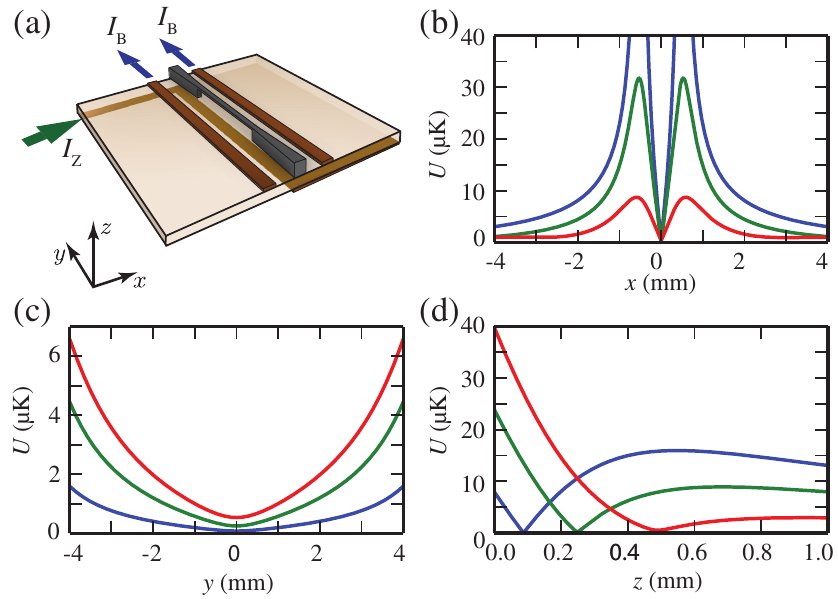}
\caption{(a) Schematic atom chip design, with grey nanostring. Wires at the bottom form a Z-shaped structure with current $I_Z$ for magnetic trapping. Bias wires at the top are placed close to the nanostring to aid magnetic transport and to cancel the nanostring's offset, $B_{0x}$.  (b,c,d) Magnetic trap cross-section for $\ket{F=1, m_F=-1} $ $^{87}$Rb atoms in three spatial directions during magnetic transport towards the $z= 0$ surface, from $I_Z = 10 A$ (red) to $6$~A (green) to $2$~A(blue), all with parallel bias wire current  $I_{B} = -5$~A.} \label{fig:design}
\end{center}
\end{figure}

Finally, we consider an architecture optimized for hybridizing ultracold atoms and a nanostring resonator, including the atoms' location and overall trapping potential; nanostring geometry; and fabrication considerations.
For atom trapping, we consider two different technologies.  The most integrated method is to use an atom-chip~\cite{reichel2011atom}. A Z-shaped wire deposited on the chip's bottom surface (Fig.~\ref{fig:design})  magnetically traps  atoms a distance 0.5~mm  from the top surface of a 0.5~mm thick chip, with $I_Z = 10$~A and equal  currents $I_B = 5$~A in two parallel wires separated by $d = 1$~mm that create a bias field. Atoms can be brought closer to the nanostring by ramping the trapping current $I_Z$ down to 2~A (Fig.~\ref{fig:design}).  
While magnetic traps are the most convenient for microfabricated integration, they permit only magnetically trappable atoms ($m_F g_F > 0$); an  optical dipole trap~\cite{Grimm:380296} is necessary for other states.  The integration of atom chips and optical traps has been done, both by maintaining a sufficient distance from the chip surface~{\cite{LeBlanc:2010fa}, or including reflective elements on the chip surface~\cite{Straatsma:2015hj} or waveguided optical fields~\cite{Burke:2002ep}.

High-tensile stress SiN nanostring resonators are a promising technology for these systems, where the string's fundamental mode $\omega_{\rm m}$ can be matched to atomic resonance $\omega_{\rm a}$ by tuning the Zeeman splitting. As an example, a 200~$\mu$m long-string,  with  width of 2.75~$\mu$m and  thickness of 350~$\mu$m has effective mass $m_{\rm eff} = 8.4\times10^{-13}$~kg and  room-temperature $Q$ factor $1.6\times 10^5$ for the fundamental mode $\omega_{\rm m}/2\pi =  850$~kHz~\cite{Biswas:2012dba}.  For moderately larger frequencies $\omega_{\rm m}$, shorter strings can be fabricated without  dramatically affecting $Q$. 
 Estimating the decoherence rate $\gamma_{\rm dec} = k_{\rm B} T_{\rm s}/\hbar Q$ (where  $T_{\rm s}$ is the substrate temperature)~\cite{Hunger:2011eo}, the room-temperature decoherence of these nanostrings will be approximately $\gamma_{\rm dec}/2\pi  = 39$~MHz, whereas at 4~K, this improves to $\gamma_{\rm dec, 4K}/2\pi  = 520$~kHz and at 10~mK, $\gamma_{\rm dec, 10mK}/2\pi  = 1.3$~kHz.  We expect improvement over the type of cantilever design previously implemented with a hot vapour of atoms ($Q = 10^3$) \cite{Wang:2006dd} and with trapped, laser-cooled atoms ($Q = 10^4$ at $r_0 = 100~\mu{\rm m}$~\cite{Montoya:2015el}.) The performance of these nanostrings is similar carbon nanotubes that use surface effects as the coupling mechanism between the mechanical motion of the atoms and the resonator~\cite{Fermani:2007di,Petrov:2009iw}, with $Q= 2.5\times10^5$, and $\gamma_{\rm dec}/2\pi  = 2.2$~Hz at a distance 1.67~$\mu$m, giving a collective coupling $\Omega = 710$~s$^{-1}$~\cite{Darazs:2014gd}.

In conclusion, the magnetic coupling in a hybrid quantum systems of ultracold atoms and magnetic mechanical resonators provides a means for quantum control through adiabatic magnetic field sweeps.  With the advantages of large $Q$-factors in SiN nanostring resonators, the controlled transfer of  quanta between  atomic and mechanical systems  provides a new pathway to mechanical cooling, a new technique for determining the temperature of mechanical resonators near their ground states, and the   entanglement of  mesoscopic objects.  As work progresses towards understanding and harnessing quantum mechanics on meso- and macroscopic scales in mechanical systems, the relative ease with which quantum degrees of freedom in atomic systems can be controlled  lends many advantages for manipulating and measuring the mechanical degrees of freedom, as demonstrated here.  Together, these hybrid systems will provide a path towards a fundamental understanding of the quantum-to-classical transition, and, especially with  cryogenic cooling and superconducting elements~\cite{Bernon:2013hm}, will offer possibilities for exploiting the richness of quantum coherence in practical technologies.

\begin{acknowledgements}
The authors gratefully acknowledge E.~Saglamyurek, J.~P.~Davis, T.~S.~Biswas, and B.~D.~Hauer for helpful conversations and ES and JPD for careful readings of this manuscript.  This work was generously supported by the University of Alberta; the Faculty of Science; the Natural Sciences and Engineering Research Council (NSERC RGPIN-2014-6618); Alberta Innovates - Technology Futures (AITF); and the Canada Research Chairs (CRC) Program.
\end{acknowledgements}


\clearpage 

\newpage

\begin{widetext}
\begin{centering}
{\large\bf Supplement to: Magnetic-field-mediated coupling and control for hybrid atomic-nanomechanical  systems}\\
\vspace{4pt}
{ A.~Tretiakov$^1$, L.~J.~LeBlanc$^{1,2}$}\\
\vspace{10pt}
{\small $^1$Department of Physics, University of Alberta, Edmonton AB, Canada}\\
{\small $^2$Canadian Institute for Advanced Research, Toronto, ON, Canada}\\
\end{centering}
\end{widetext}

\renewcommand{\thesubsection}{S\arabic{subsection}}   
\renewcommand{\thetable}{S\arabic{table}}   
\renewcommand{\theequation}{S\arabic{equation}}
\renewcommand{\thefigure}{S\arabic{figure}}
\setcounter{figure}{0}
\setcounter{equation}{0}

\subsection{Temperature probe calculations}

To determine the spin-population dependence for the temperature probe application we consider the occupation of phonon modes of the mechanical oscillator at various temperatures, as determined by the Bose-Einstein statistics for a thermal distribution
\begin{align}
P_\omega(n) = \frac{e^{-n\hbar\omega_{\rm m}/k_B T}}{\sum_n e^{-n\hbar\omega_{\rm m}/k_BT}} = e^{-n\hbar\omega_{\rm m}/k_B T}\left(1-e^{-\hbar\omega_{\rm m}/k_B T}\right).\nonumber
\end{align}

Next, we consider the action of the Landau-Zener sweeps.  For $n>0$:
\begin{align}
\ket{n,\uparrow} &\rightarrow \Big[\sqrt{p_{\rm LZ}(n+1)}\ket{n+1,\downarrow}+\nonumber\\
& \qquad \qquad\qquad \qquad \sqrt{1-p_{\rm LZ}(n+1)}~e^{i\phi}\ket{n,\uparrow}\Big] \nonumber \\
\ket{n,\downarrow}& \rightarrow \left[\sqrt{p_{\rm LZ}(n)}\ket{n-1,\uparrow}+\sqrt{1-p_{\rm LZ}(n)}~e^{-i\phi}\ket{n,\downarrow}\right]\nonumber
\end{align}
and for $n = 0$
\begin{align}
\ket{0,\uparrow} &\rightarrow \left[\sqrt{p_{\rm LZ}(1)}\ket{1,\downarrow}+\sqrt{1-p_{\rm LZ}(1)}~e^{i\phi}\ket{0,\uparrow}\right]\nonumber \\
\ket{0,\downarrow}& \rightarrow \ket{0,\downarrow}\nonumber
\end{align}
where $p_{\rm LZ}(n)$ is the phonon-dependent Landau-Zener (LZ) probability of making an adiabatic transition across the  level crossing.  The phonon dependence comes from the amplitude dependence of coupling strength, such that the avoided crossings have a separation $\sqrt{n}g_0$ in any Jaynes-Cummings ladder.   The calculated phonon-number-dependence of the probability for adiabatic Landau-Zener transitions is shown in Fig.~\ref{fig:LZprob}.

\begin{figure}[t!]  
\begin{center}
\includegraphics[width = 85mm]{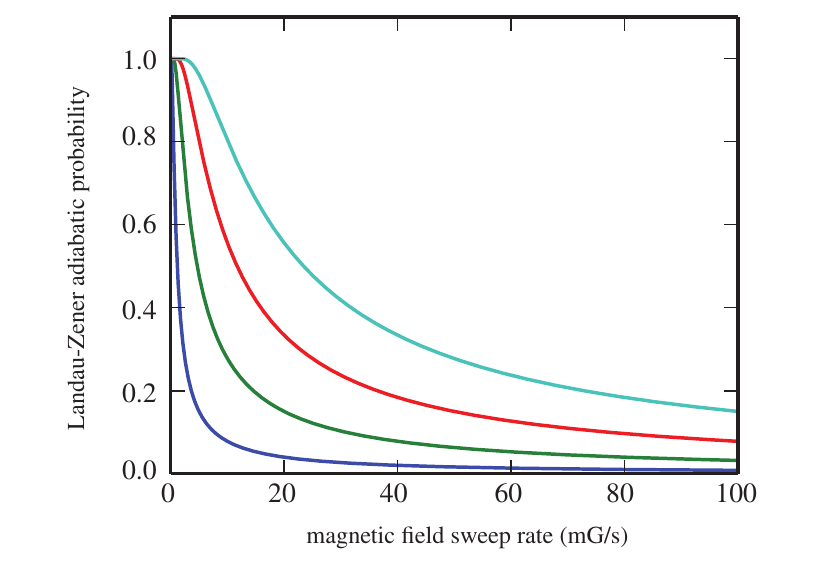}
\caption{(a) Calculated Landau-Zener probability for adiabatic transition for different average phonon numbers $\bar n$ vs.\ rate of the sweep $\Delta B/\Delta T$.  Blue: $n = 5$, green: $n = 20$, red: $n = 50$, cyan: $n = 100$.}
\label{fig:LZprob}
\end{center}
\end{figure}

To determine the probability of spin flip after a sweep from one state to another, we determine the probability of occupation for each phonon number state for a thermal distribution at temperature $T$, and then apply the LZ sweep to a single state of this form.  These states are numerically represented and include phonon number states up to 40 times their average phonon number in the basis $\ket{\psi} \rightarrow \big[ \sqrt{P_\omega(0)}\downarrow,$ $ \sqrt{P_\omega(0)}\uparrow,$ $ \sqrt{P_\omega(1)}\downarrow,$ $ \sqrt{P_\omega(1)}\uparrow,$ $ \sqrt{P_\omega(2)}\downarrow, $ $\sqrt{P_\omega(2)}\uparrow, $ $ \ldots \big]$.  The L-Z sweep can be implemented as a matrix operation on this state
\begin{align}
\hat{\rm LZ} =
 \left[\begin{array}{ccccccc}
1 & 0 & 0 & 0 & 0 & 0 &\\
0 & \tilde p(1) & \bar p(1)  & 0 & 0 & 0 &\\
0 & \bar p(1) & \tilde p(1) & 0 & 0 & 0 & \ldots\\
0 & 0 & 0 & \tilde p(2) & \bar p(2) & 0 &\\
0 & 0 & 0 & \bar p(2) & \tilde p(2) & 0& \\
0 & 0 & 0 & 0 & 0 & \tilde p(3) &\\
 &  &  \vdots& & & & \ddots
\end{array} \right]\nonumber
\end{align}
where $\bar p(n) = \sqrt{p_{\rm LZ}(n)}$ and $\tilde p(n) = \sqrt{1-p_{\rm LZ}(n)}$.  
 The resulting state is then analyzed in terms of the spin population to determine how many spins flipped and how many stayed the same by finding the expectation value $\bra{\psi_{\rm f}}\hat{\rm S}_\uparrow \ket{\psi_{\rm f}}$ if the originally atom was in spin-down state and $\bra{\psi_{\rm f}}\hat{\rm S}_\downarrow \ket{\psi_{\rm f}}$ if the atom was with spin up, where $\ket{\psi_{\rm f}}$ is the wave function after the sweep and the spin operators are given by
 \begin{align}
 \hat{\rm S}_\downarrow =
 \left[\begin{array}{cccccc}
1 & 0 & 0 & 0 &0 \\
0 & 0 & 0 & 0 &0 \\
0 & 0 & 1 & 0 &0 &\ldots \\
0 & 0 & 0 & 0 &0 \\
0 & 0 & 0 & 0 &1 \\
& & \vdots & & & \ddots
\end{array} \right]\nonumber
\qquad 
 \hat{\rm S}_\uparrow =
 \left[\begin{array}{cccccc}
0 & 0 & 0 & 0 &0 \\
0 & 1 & 0 & 0 &0 \\
0 & 0 & 0 & 0 &0 &\ldots \\
0 & 0 & 0 & 1 &0 \\
0 & 0 & 0 & 0 &0 \\
& & \vdots & & & \ddots
\end{array} \right].
 \end{align}
 
Using this formalism, we can determine the probability for spin state transfer (``spin flipping'') for a particular  distribution of energies among the phonon number states, and the ratio of these probabilities for different initial states. While the above methodology will work for any distribution, we consider here the Bose-Einstein thermal distribution of phonon modes, for which we can also obtain analytic expressions.  Starting in the atomic spin state  $\ket{\psi_{i,at}} = \ket{\uparrow}$, the a measurement of the final spin expectation value is $p_{\rm flip, \downarrow} = \bra{\psi_{\rm f}}\hat{\rm S}_\uparrow \ket{\psi_{\rm f}}= \sum_{n=0} p_{\rm LZ}(n+1) x^n(1-x)$, and starting with the atomic spin state  $\ket{\psi_{i,at}} = \ket{\uparrow}$ results in a measurement of the spin-flipping probability $p_{\rm flip, \uparrow} = \bra{\psi_{\rm f}}\hat{\rm S}_\downarrow \ket{\psi_{\rm f}}= \sum_{n=1} p_{\rm LZ}(n) x^n(1-x)$, where $x = \exp(-\hbar \omega_{\rm m}/k_B T)$.  The ratio of these probabilities is simply the Boltzmann factor for the zero-phonon state $p_{\rm flip \downarrow}/ p_{\rm flip \uparrow}= \exp(-\hbar \omega_{\rm m}/k_B T)$, and a measure of the ratio of spins flipped gives a measure of the ground state occupation, and thus, a measure of temperature.

For estimates of temperature at higher phonon number occupations, one could use the amplitude dependence of the coupling parameter together with the  LZ spin-flip probability's exponential dependence on the coupling parameter.  By measuring the probability of a spin flip, extracting the coupling parameter, and relating this to the amplitude, the temperature can be found through the relationship
\begin{align}
\langle \hat\alpha^2 \rangle =  \frac{\hbar}{2m_{\rm eff} \omega_{\rm m}} \left( \frac{1+ e^{-\hbar \omega_{\rm m}/k_B T}}{1- e^{-\hbar \omega_{\rm m}/k_B T}}\right).\nonumber
\end{align}
where the nanostring's displacement operator is  $\hat\alpha =  \alpha_0(\hat a +  \hat a^\dagger)$.   
\begin{figure}[t!]  
\begin{center}
\includegraphics[]{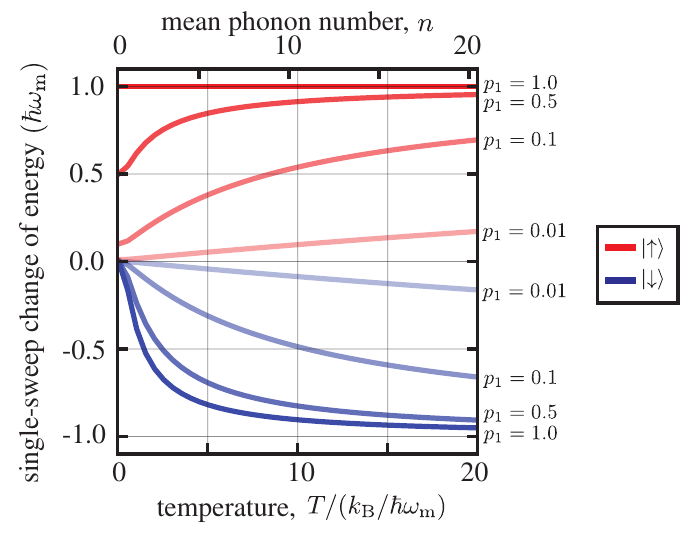}
\caption{(a) Calculated energy change per Landau-Zener sweep, for initial states $\ket{\psi_{\rm init}} = \ket{\uparrow}$ (red) and  $\ket{\psi_{\rm init}} = \ket{\downarrow}$ (blue).  Calculation are performed for sweeps where the lowest Landau-Zener probability is $p_{\rm LZ}(n=1) =p_1 =  $ 1.0, 0.5, 0.1, and 0.01.}
\label{fig:energyChange}
\end{center}
\end{figure}

The temperature dependence of the cooling scheme also depends on the occupation of the ground phonon state (which is not active in cooling).  Fig.~\ref{fig:energyChange} shows the change in energy for a single sweep starting in either the  atomic spin state  $\ket{\psi_{i,at}} = \ket{\uparrow}$ (red) or $\ket{\psi_{i,at}} = \ket{\downarrow}$ (blue) for several different $n_{\rm ph} = 1$ LZ probabilities for high-to-low magnetic field sweeps.

\subsection*{Calculating effective cooling}

The calculation of effective cooling (as measured by the reduction of $\bar n$) in Fig.~2d is found by assuming the above dynamics of the LZ sweeps, and by simulating optical pumping as an incoherent process that transfers populations (as opposed to amplitudes, as above).  After a series of one-LZ-sweep, one-OP-pulse steps, $\bar n$ is determined, and this is plotted.

\subsection{Limits on coupling strength}

\begin{figure}[t]
	\begin{center}
		\includegraphics[width=85mm]{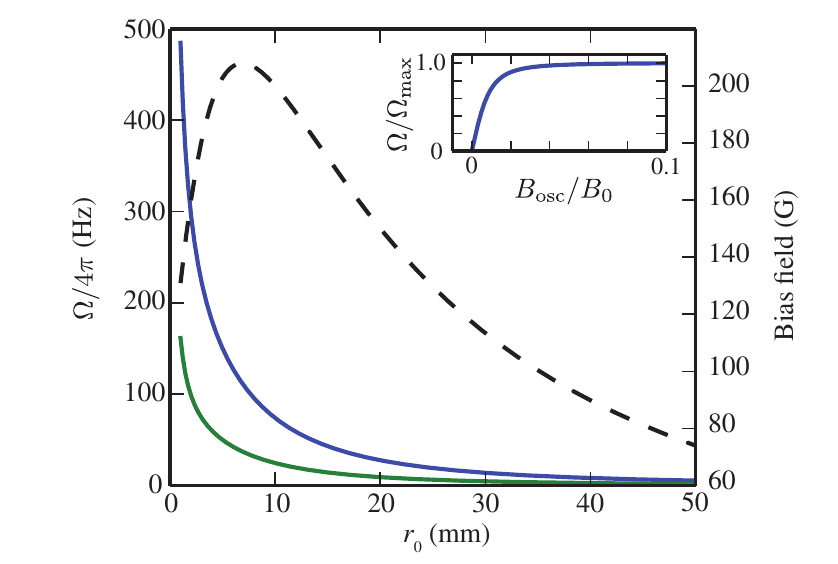}
	\end{center}
	\caption{Local bias field (dashed, right axis) produced by two long on-chip wires (width of 10~$\mu$m, thickness $4~\mu$m, separated by 10~$\mu$m) carrying a current of 1 A, and corresponding coupling parameters for a nanostring with permanent magnet (blue) and dc current (green), where the permanent magnet's moment and current are chosen to create constant offset equal to the corresponding bias field. The nanostring is located at $4~\mu$m above the chip's surface.  Inset: the ratio of coupling constant $\Omega$ to the maximum possible coupling constant $\Omega_{\rm max} = \Omega B_0 / \chi B_{\rm osc}$, as a function of oscillating-field-to-bias-field ratio, showing a maximum value when the oscillating field magnitude is large.  For a fixed available $B_0$, this sets a limit for the coupling strength.}
	\label{fig:Bias}
\end{figure}

As described in the main text, both the permanent-magnet and direct-current coupling schemes result in an inconvenient stationary magnetic field $B_{0x}$ ($B_{0x, {\rm d}}$ or $B_{0x, I}$). Estimates of the coupling parameter above (Eq.~2) assume  the external quantization field $B_{0z}$ is perpendicular to the oscillating field $B_{\rm m,0}$, but the stationary component of the resonator's field is parallel.  Since the relevant quantization field is the quadrature sum of these components, $\Omega$  is reduced because the oscillating  field orthogonal to the effective quantization axis is
$B_{\rm m,\perp} = {B_{0z}B_{\rm m,0}}/[B_{0z}^2 + \chi^2 B_{\rm m,0}^2]^{1/2}$,
where the parameter fixing the ratio of oscillating to constant field is $\chi = r_0/3\alpha$ or $\chi = r_0/\alpha$, in the case of the permanent-magnet or the direct-current nanostrings, respectively [Eqs.~(3) and (4)]. In an experiment where the external field is technically limited to  $B_{0z}$, attempts to increase the coupling strength (e.g. by putting more magnetic material in one case and running larger current in the other) shift  the quantization axis towards $\ex$ and, the coupling parameter is limited to $\hbar\Omega_{\rm max} = (  g_F \muB \Omega B_{0z}/\hbar\chi )\bra{F^\prime m_F^\prime} \hat{F}_x \sin\theta+\hat{F}_z \cos\theta \ket{F m_F}$.
One solution to this unwanted bias field is to use a compensating field created by permanent magnetic elements~\cite{TreutleinPRL2007}[49] or in a simpler, tunable design, with two parallel wires on a chip surface to create bias fields (see Fig.~\ref{fig:Bias}).
While the ac-current method does not suffer from this accompanying bias field, this current will drive oscillation in the resonator~\cite{Biswas:2012dba} and prevent attempts to reach the lowest temperatures.
\vspace{12pt}
\hrule
\vspace{12pt}
[49] Y.~F. Leung {\it et~al.}, Rev Sci Instrum {\bf 85},  053102  (2014).

\end{document}